\documentclass[fleqn,usenatbib]{mnras}

\usepackage{newtxtext,newtxmath}
\usepackage[T1]{fontenc}

\DeclareRobustCommand{\VAN}[3]{#2}
\let\VANthebibliography\thebibliography
\def\thebibliography{\DeclareRobustCommand{\VAN}[3]{##3}\VANthebibliography}

\usepackage{graphicx}	% Including figure files
\usepackage{amsmath}	% Advanced maths commands
\title[Dense gas in M31 GMCs]{Tracing Dense Gas in Six Resolved GMCs of the Andromeda Galaxy}

\author[J. Forbrich et al.]{
Jan Forbrich,$^{1}$\thanks{E-mail: j.forbrich@herts.ac.uk (JF)}
Charles J. Lada,$^{2}$
J\'er\^ome Pety,$^{3}$
and Glen Petitpas$^{4}$
\\
$^{1}$University of Hertfordshire, Centre for Astrophysics Research, College Lane, Hatfield, AL10 9AB, UK\\
$^{2}$Center for Astrophysics | Harvard \& Smithsonian, 60 Garden St, Cambridge, MA 02138, USA\\
$^{3}$Institut de Radio Astronomie Millim\'etrique, 300 rue de la Piscine, 38406 Saint-Martin-d'H\`eres, France\\
$^{4}$MIT Kavli Institute for Astrophysics and Space Research, Massachusetts Institute of Technology, 77 Massachusetts Avenue, Cambridge, MA 02139, USA
}

\date{Accepted 2023 August 23. Received 2023 August 21; in original form 2023 May 12}
\pubyear{2023}

\begin{document}
\label{firstpage}
\pagerange{\pageref{firstpage}--\pageref{lastpage}}
\maketitle

\begin{abstract}
%This is a simple template for authors to write new MNRAS papers. The abstract should briefly describe the aims, methods, and main results of the paper. It should be a single paragraph not more than 250 words (200 words for Letters). No references should appear in the abstract.
We present dense-gas--tracing molecular observations of six resolved Giant Molecular Clouds (GMCs) in the Andromeda Galaxy (M31). Using the NOEMA interferometer, we observed the transitions of HCN(1-0), HCO$^+$(1-0), and HNC(1-0), as well as $^{13}$CO(1-0) and 100~GHz continuum emission. This complements our earlier work with the Submillimeter Array (SMA), including resolved dust continuum detections of these clouds at 230~GHz. In this work, we first compare different continuum measurements to conclude that the average free-free contamination of the observed flux is 71\% at 3~mm but only 13\% at 1~mm, confirming that emission at 3~mm is less reliable than that at 1~mm for calculating dust masses of star-forming clouds. While the $^{13}$CO emission is more extended than both HCN and HCO$^+$ emission, which in turn is more extended than HNC emission, we find that both HCN and HCO$^+$ are spatially coincident with, and similarly extended as, the 230~GHz dust emission. This suggests that both the 230~GHz dust continuum and most importantly the HCN emission traces the dense gas component of these GMCs. From comparison of the molecular emission with dust masses derived from the 230~GHz continuum emission, we obtain the first direct measurements of the dust-mass-to-light ratios ($\alpha^\prime_{HCN}$ and $\alpha^\prime_{HCO^+}$) in GMCs of an external galaxy. For HCN, the result is broadly similar to a measurement in the local Perseus cloud suggesting that these are indeed dense gas conversion factors. A larger cloud sample will be required to assess whether HCN is tracing comparable cloud-scale density regimes across the environments of M31.
\end{abstract}


%%%%%%%%%%%%%%%%%%%%%%%%%%%%%%%%%%%%%%%%%%%%%%%%%%

%%%%%%%%%%%%%%%%% BODY OF PAPER %%%%%%%%%%%%%%%%%%

\section{Introduction}\label{sec:int}

Because of their relatively high dipole moments compared to CO, molecules such as HCN, CS, NH$_3$, N$_2$H$^+$ and HCO$^+$, have long been regarded as tracers of the dense gas (i.e., n(H$_2$) $\gtrsim$ 10$^4$ cm$^{-3}$) component of molecular clouds. 
In the Milky Way such dense gas has been long known to be intimately associated with the star formation process (e.g., \citealp{lad74,mye87,lad92}). In external galaxies this connection was perhaps best illustrated in the well known study of \citet{gas04} who reported a significant correlation between the far-infrared (FIR) and HCN(1-0) luminosities in a large sample of star forming galaxies. Since the FIR luminosity is a proxy for the global star formation rate (SFR) and HCN luminosity a proxy for the mass of dense molecular gas (M$_{dg}$) in a galaxy, this finding indicated the existence of a linear star formation law for galaxies, connecting the global SFR with the mass of dense molecular gas within them. Subsequent observations were able to extend the Gao-Solomon relation and the  linear `dense gas' star formation law to clouds in the Milky Way \citep{wue05,lad12,jim19}. 

Studies of local Milky Way clouds using infrared dust extinction measurements rather than molecular line emission to trace the molecular mass showed that the SFR was more tightly correlated with the mass of gas at high column density (A$_K$ $\gtrsim$ 0.8 magnitudes) than with the total cloud mass  (i.e., A$_K$ $>$ 0.1 magnitudes; \citealp{lad10}).
Because of the stratified, filamentary geometry of molecular clouds in the Galaxy, the higher dust column densities (i.e., A$_K$ $\gtrsim$ 0.8 mag) typically correspond to the densest (i.e., n(H$_2$) $>$ 10$^4$ cm$^{-3}$) regions of the GMCs. This is supported by numerous lines of empirical and theoretical evidence including recent 3D extinction maps \citep{zuc21} of local clouds, models of a cloud's radial structure  \citep{ber01} and numerical simulations of cloud formation and evolution \citep{bis19}. Thus the Galactic star formation law derived from extinction measurements was in apparent agreement with the extragalactic relation derived by Gao \& Solomon from HCN measurements. 

Due to complications in interpreting the HCN observations, estimating the actual densities required for its detection remained uncomfortably uncertain \citep{eva20}. In the case of HCN emission, this is because it is often optically thick, and radiative trapping can reduce the density required to excite the molecule to detectable levels. At $T_k = 20$~K, the critical density required for an optically thin HCN(1-0) line to be detectable is $n_{\rm crit}({\rm HCN}) = 3\times10^5$~cm$^{-3}$, while radiative trapping can reduce the effective excitation density for detection to $n_{\rm eff}({\rm HCN}) = 4.5\times10^3$~cm$^{-3}$, if the line is optically thick \citep{shi15}. Indeed, recent mapping studies of HCN in nearby Galactic molecular clouds found the HCN emission to be extended beyond the boundaries of the high extinction--high density regions traced by dust in the clouds \citep{pet17,shi17,kau17,bar20,dal23}. 
While observations of entire local GMCs are still rare,
these observations have cast some doubt on the use of HCN and similar molecules to trace dense gas and therefore on the nature of the star formation law derived from the Gao-Solomon relation.  

However, a recent and detailed study of the Perseus cloud, a local Milky Way GMC, now suggests that molecules like HCN are in fact good dense gas tracers and with proper calibration can even be used to trace the densest gas defined by dust extinction measurements within molecular clouds\footnote{Here we follow the convention that dense gas is defined to be gas above the A$_K$ $=$ 0.8 (or A$_V$ = 7.3) magnitude threshold.}. Specifically,  \citet{dal23} obtained the first deep and spatially complete survey of HCN emission from an individual molecular cloud (the Perseus cloud) and combined the survey with existing and complete infrared dust extinction maps \citep{zar16} to investigate the relation between HCN and dust column density. Similar to earlier studies they found HCN emission to be extended beyond the extinction identified high density regions of the cloud with 60\% of the emission originating in regions below 0.8 magnitudes of infrared extinction. Although radiative transfer modelling indicated that the HCN emission arose from gas with densities ranging from $\sim$ 200-35000 cm$^{-3}$, they found the average volume density in the HCN emitting gas to be 10$^4$ cm$^{-3}$, 
in line with expectations for the effective excitation density for optically thick emission and indicating that HCN was still reasonably tracing high density gas. To produce a dense gas calibrated $\alpha_{HCN}$ 
Dame \& Lada used their data to measure the total HCN luminosity of the cloud and the infrared extinction maps to measure the dense gas mass. Combining the two measurements they were able to produce a direct, empirical derivation of the dense gas conversion factor, $\alpha_{\rm HCN}$ $=$ $M_{dg}\over{L_{HCN}}$ $=$ 92 M$_\odot$ (K~km s$^{-1}$pc$^2$)$^{-1}$, considerably higher than the generally adopted value of 10 \citep{gas04}. Whether or not these Perseus results are characteristic of molecular clouds in general remains to be determined.

Complete maps of individual molecular clouds in HCN emission, either in our own Galaxy or in  external galaxies, are still rare making the few existing determinations of $\alpha_{HCN}$ in the Milky Way, based on incomplete mapping observations, (e.g., \citealp{pet17,shi17,kau17}) uncertain. 
There are no direct measurements of $\alpha_{HCN}$ in individual GMCs within external galaxies. Because the knowledge of the dense gas contents of molecular clouds are so important for star formation studies, direct measurements of $\alpha_{\rm HCN}$ in molecular clouds of an external galaxy are highly desirable. Although sufficiently deep observations of HCN are possible with modern interferometers, direct determinations of $\alpha_{HCN}$ also require reliable measurements of the dense gas masses and these are best obtained from observations of the dust that, up to now, were only possible in the nearby clouds of the Milky Way. 
With the recent deployment of wide-band continuum receivers on the Submillimeter Array (SMA), resolved interferometric measurements of dust emission have been obtained for the first time from individual GMCs in the nearest spiral galaxy, M31 \citep{for20,via21}. As a result, measurements of  dust derived masses are now available for a significant sample of clouds in M31. Acquisition of high resolution HCN observations of these clouds would enable both the first direct measurements of $\alpha_{\rm HCN}$ in an external galaxy as well as a determination of the extent of cloud to cloud variations in the HCN conversion factor.

In the following paper we report the results of a pilot study of HCN(1-0), HCO$^+$(1-0), HNC(1-0) and continuum emission in a sample of six resolved GMCs in the Andromeda Galaxy (M31) obtained with the NOEMA telescope at $\sim$ 100 GHz. Since, as we show in this paper, the 100~GHz continuum is contaminated by free-free emission, we combine our NOEMA observations with the SMA continuum observations at 230~GHz to make the first direct measurements of $\alpha^\prime_{\rm HCN}$ and $\alpha^\prime_{\rm HCO^+}$ toward molecular clouds in an external galaxy.

In section~\ref{sec:obs} we describe the NOEMA and SMA observations. In section~\ref{sec:res} we report the results of both the molecular line and continuum observations. In section~\ref{sec:ana} we analyze the data and report our determinations of the dense gas conversion factors, $\alpha_{\rm HCN}$ and $\alpha_{\rm HCO^+}$. We also separately discuss comparisons of the HCN and HCO$^+$ observations and the HCN and HNC observations. In section~\ref{sec:sum} we summarize the results of the paper.

%%%%%%%%%%%%%%%%%%%%%%%%%%%%%%%%%%%%%%%%%%%%%%%%%%%%%%%%%

\section{Observations} \label{sec:obs}

\subsection{Sample selection}

The target GMCs selected for study here are a subset of a much larger  population of GMCs identified and mapped in an earlier Submillimeter Array (SMA) survey of 230~GHz continuum and CO emission across M31. The targets for the SMA survey were drawn from the {\it Herschel} survey of Giant Molecular Associations (GMAs) in M31 by \citet{Kirk2015}. For this study we selected a handful of the brightest 230-GHz continuum sources detected in the SMA experiment \citep{for20,via21}. Besides increasing the likelihood of detecting HCN emission in these clouds, this selection strategy also allowed for an assessment of the complementary ability to detect 100-GHz continuum emission with NOEMA. 

Five of the six selected GMCs (all apart from K\,190) are associated with H\,{\sc ii} regions as determined by the presence of 10 GHz VLA continuum sources within their boundaries. The nominal radio flux density of these candidate H\,{\sc ii} regions ranges from twice to twenty times that of the Orion Nebula (K176 and K191, respectively), where our experiment was designed to detect Orion-like H\,{\sc ii} regions (Toomey et al. {\it in prep.}).

\subsection{NOEMA observations}

The six GMCs from the SMA sample were observed with NOEMA in the 3~mm band and in track-sharing mode, in programs S20AX (K026, K191, and K213) and W20BN (K136, K176, and K190). The spectral setup included a 3~mm continuum band, as well as spectral lines HCN(1-0), HCO$^+$(1-0), HNC(1-0), and $^{13}$CO(1-0). In S20AX,  a total of 19 hours of overall telescope time was collected on three separate occasions in July and August 2020, with weather conditions ranging from good to excellent. Observations for W20BN yielded 17.5 hours of overall telescope time in May 2021, collected on four separate occasions with weather conditions ranging from poor to excellent. Spectral windows ranging in width from 192 to 320~MHz were set up for the targeted spectral lines, at a resolution of 62.5~kHz, and the continuum was covered in the full band (16~GHz times two polarizations) at a resolution of 2~MHz. 

Calibration (bandpass, phase, amplitude, and flux) was done using standard methods implemented in GILDAS/CLIC pipeline. Imaging and deconvolution used standard methods implemented in GILDAS/MAPPING. For the continuum, we first filtered 30 MHz around each of the expected line frequencies before imaging it.  We computed the LSB and USB fluxes inside ellipses designed to cover the detected emission and we merged the LSB and USB $(u,v)$ tables with a correction for the spectral index before re-imaging the combined continuum $(u,v)$ table, even though the spectral index itself cannot be reliably determined from the two sidebands due to the limited frequency baseline for this purpose. Due to the non-detection of K190 and indications of different spectral indices in K213, these targets were separately imaged in the two sidebands.

For lines, we resampled the spectra to a velocity grid of 200 channels of 1 km\,s$^{-1}$ spacing and centered on the line rest frequency at the systemic velocity of the targeted cloud. We then subtracted a 0th order baseline out of each visibility, obtained outside of a window of 20 MHz around the line rest frequency. We imaged and deconvolved the data without support first. We then defined a large circular support of 25$''$ radius to improve the CLEANing convergence. The data was then converted from Jy/beam to K. We experimented with a primary beam correction but found that for the targets concerned the correction was limited to $<5$\%, below other sources of uncertainty, including the knowledge of the primary beam shape. The images thus have not been corrected for the primary beam shape.

\subsection{SMA observations}

Here we only make use of the 230~GHz wideband continuum and $^{13}$CO(2-1) maps obtained in our SMA program toward the NOEMA targets. Details regarding the SMA observations were previously reported in  \citet{for20} and \citet{via21}. Of interest here we note that the synthesized SMA beam size was  nominally very similar to that of the NOEMA observations reported here as shown in Table~\ref{tab_sample}. Each SMA measurement also consisted of a full observing track providing a good range in $(u,v)$ coverage. The HPBW primary beam was 55$''$ corresponding to $\sim$200~pc at the distance of M31, with spatial sensitivity to size scales of up to $\sim$100~pc. The spectral resolution was 140 kHz per channel and the $^{13}$CO data were binned to a velocity resolution of 1.3 km s$^{-1}$. 

\begin{figure*}
\centering
\begin{minipage}{.359\textwidth}
\includegraphics[width=\textwidth]{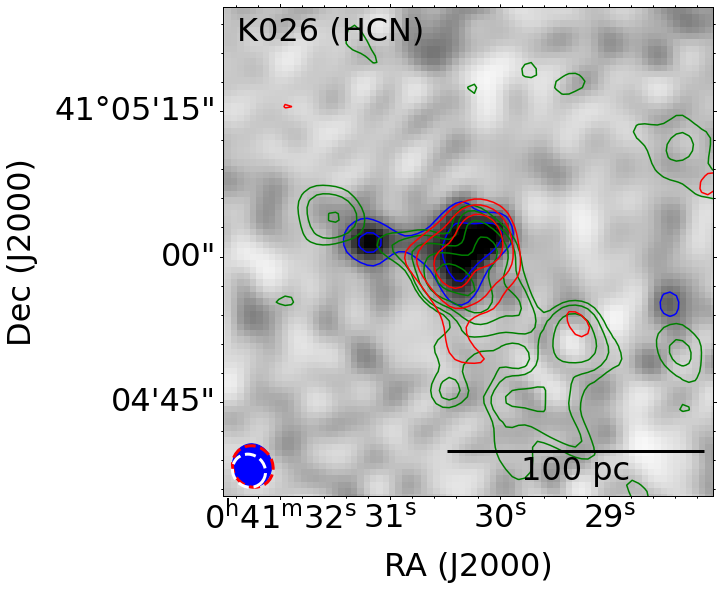}
%\centerline{K026: HCN(1-0)}
\end{minipage}
\begin{minipage}{.25\textwidth}
\vspace{-8mm}
\includegraphics[width=\textwidth]{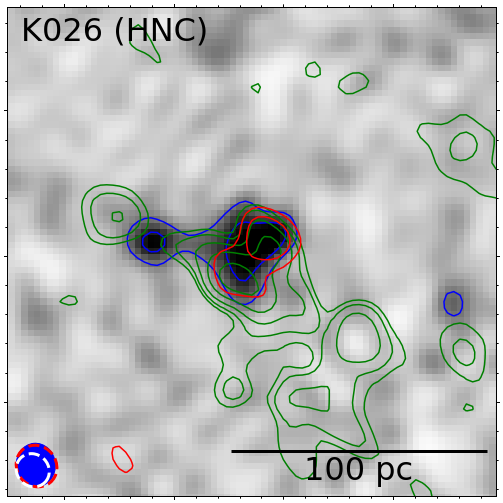}
%\centerline{HNC(1-0)}
\end{minipage}
\begin{minipage}{.25\textwidth}
\vspace{-8mm}
\includegraphics[width=\textwidth]{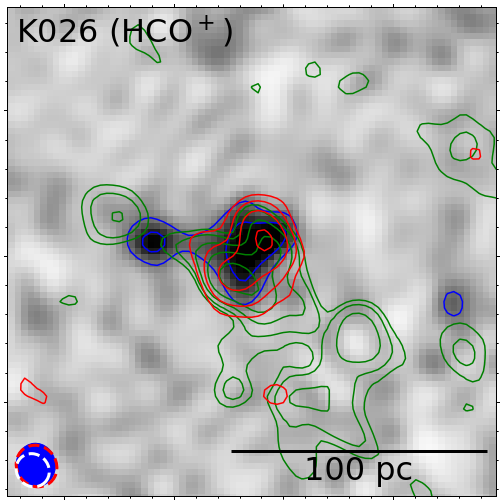}
%\centerline{HCO$^+$(1-0)}
\end{minipage}

\medskip

\begin{minipage}{.359\textwidth}
\includegraphics[width=\textwidth]{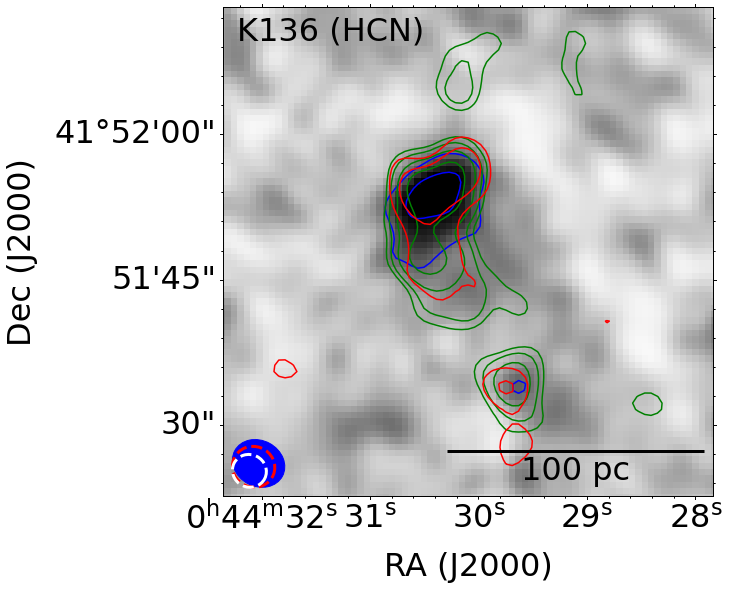}
%\centerline{K136: HCN(1-0)}
\end{minipage}
\begin{minipage}{.25\textwidth}
\vspace{-8mm}
\includegraphics[width=\textwidth]{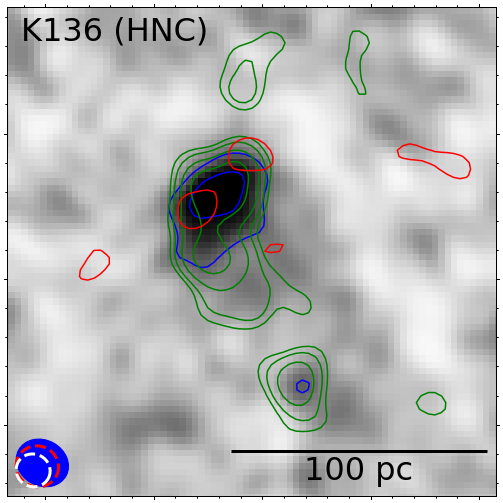}
%\centerline{HNC(1-0)}
\end{minipage}
\begin{minipage}{.25\textwidth}
\vspace{-8mm}
\includegraphics[width=\textwidth]{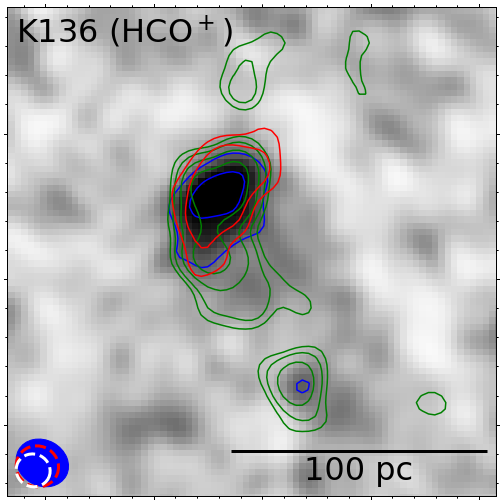}
%\centerline{HCO$^+$(1-0)}
\end{minipage}

\medskip
          
\begin{minipage}{.359\textwidth}
\includegraphics[width=\textwidth]{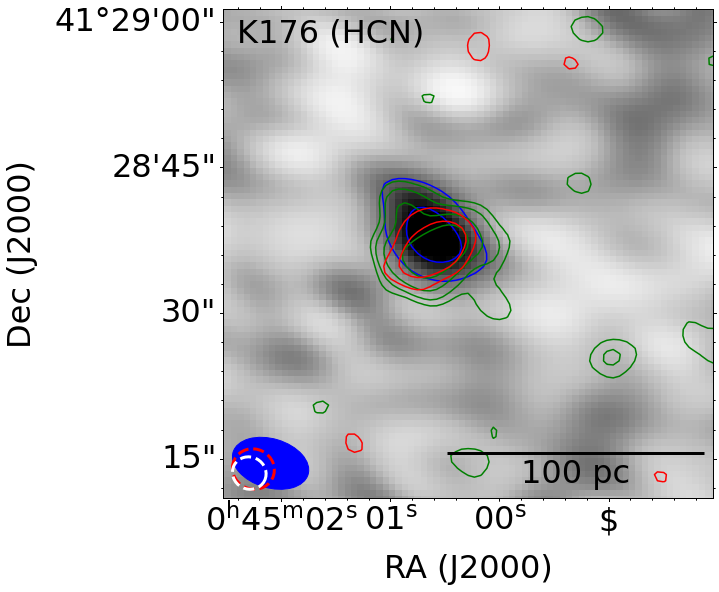}
%\centerline{K176: HCN(1-0)}
\end{minipage}
\begin{minipage}{.25\textwidth}
\vspace{-8mm}
\includegraphics[width=\textwidth]{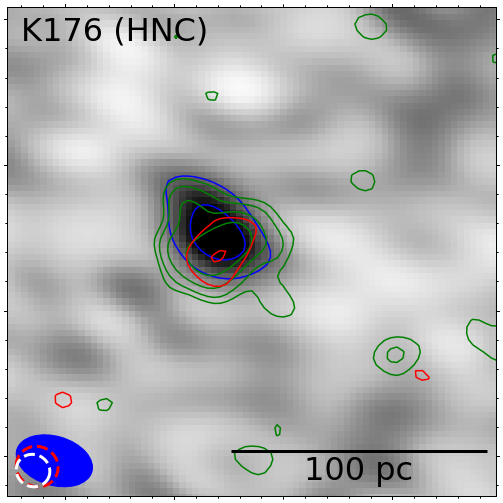}
%\centerline{HNC(1-0)}
\end{minipage}
\begin{minipage}{.25\textwidth}
\vspace{-8mm}
\includegraphics[width=\textwidth]{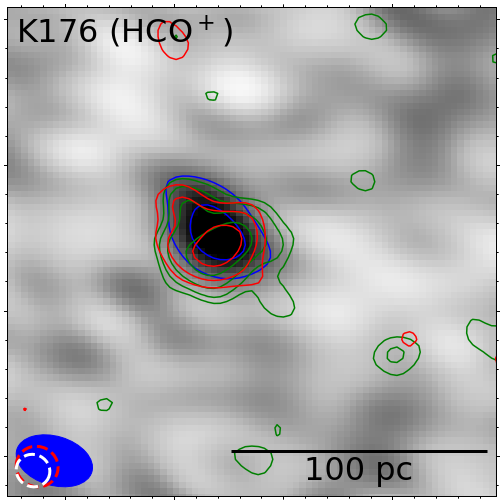}
%\centerline{HCO$^+$(1-0)}
\end{minipage}

\medskip

\begin{minipage}{.359\textwidth}
\includegraphics[width=\textwidth]{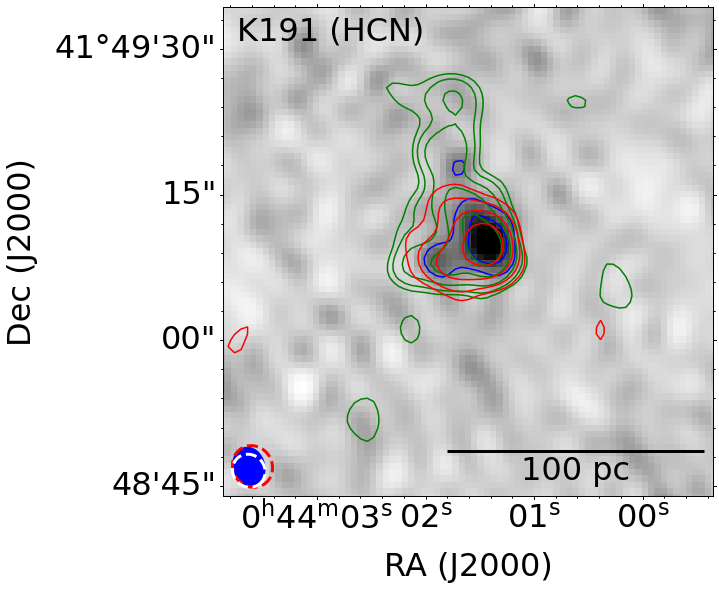}
%\centerline{K191: HCN(1-0)}
\end{minipage}
\begin{minipage}{.25\textwidth}
\vspace{-8mm}
\includegraphics[width=\textwidth]{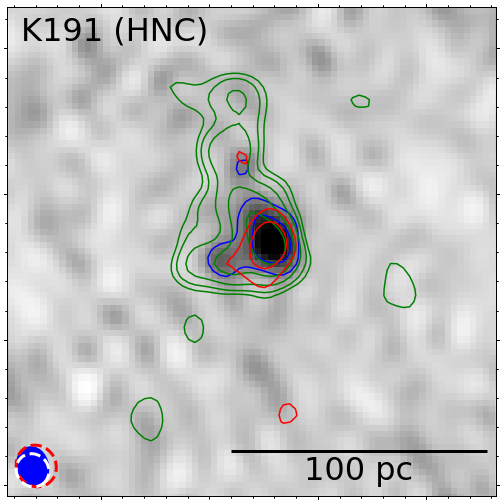}
%\centerline{HNC(1-0)}
\end{minipage}
\begin{minipage}{.25\textwidth}
\vspace{-8mm}
\includegraphics[width=\textwidth]{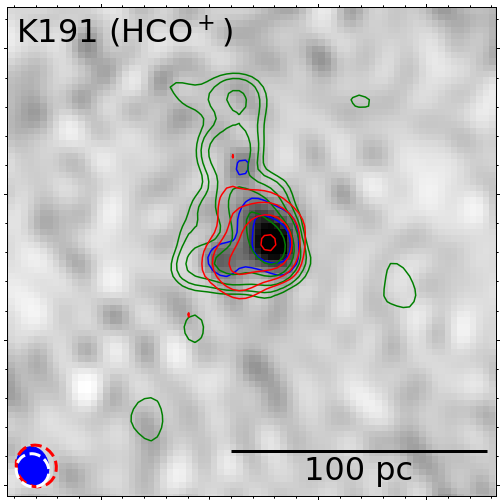}
%\centerline{HCO$^+$(1-0)}
\end{minipage}

%\medskip

%\begin{minipage}{.359\textwidth}
%\includegraphics[width=\textwidth]{m31_213_smacont_13co10_hcn10_pub_v3.png}
%%\centerline{K213: HCN(1-0)}
%\end{minipage}
%\begin{minipage}{.25\textwidth}
%\vspace{-8mm}
%\includegraphics[width=\textwidth]{m31_213_smacont_13co10_hnc10_pub_v3.png}
%%\centerline{HNC(1-0)}
%\end{minipage}
%\begin{minipage}{.25\textwidth}
%\vspace{-8mm}
%\includegraphics[width=\textwidth]{m31_213_smacont_13co10_hcop10_pub_v3.png}
%%\centerline{HCO$^+$(1-0)}
%\end{minipage}

\caption{230~GHz (SMA) dust continuum image of K026, K136, K176, K191, and K213 (top to bottom) with contour lines indicating the 3,6,12,...$\sigma$ extent of the 230~GHz (SMA) continuum (blue dashed), $^{13}$CO(1-0) (green), and (left) HCN(1-0) (red), (center) HNC(1-0) (red), (right) HCO$^+$(1-0) (red). The synthesized beam sizes are indicated in the lower left corner in blue (230~GHz dust continuum), red dashed contours (HCN, HNC, and HCO$^+$) and white dashed contours ($^{13}$CO(1-0)).
\label{fig_maps}}
\end{figure*}

\begin{figure*}
\centering
\begin{minipage}{.359\textwidth}
\includegraphics[width=\textwidth]{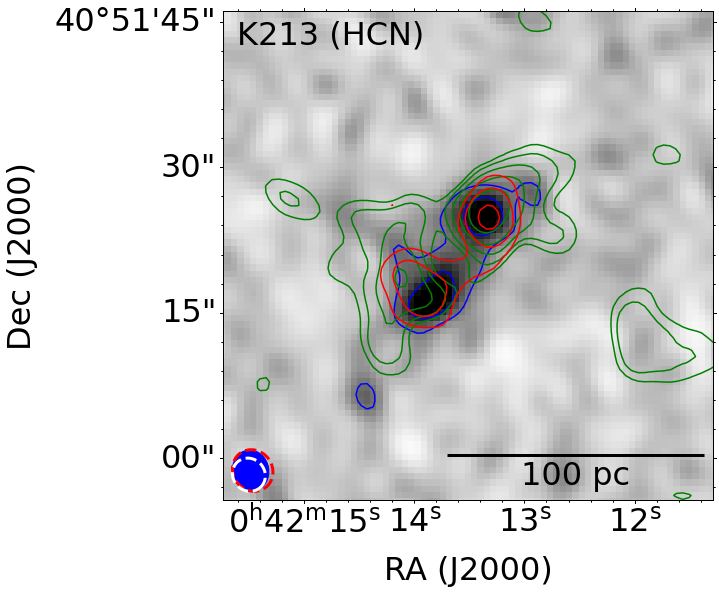}
%\centerline{K213: HCN(1-0)}
\end{minipage}
\begin{minipage}{.25\textwidth}
\vspace{-8mm}
\includegraphics[width=\textwidth]{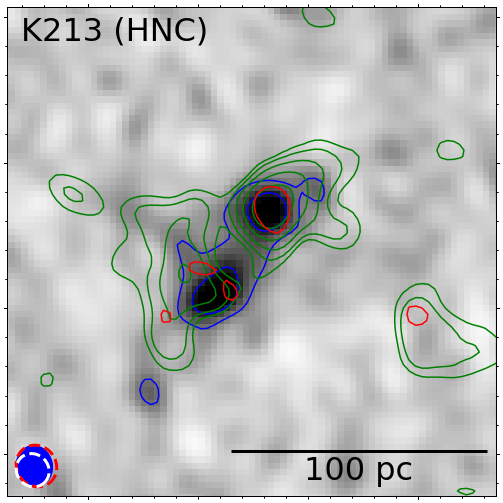}
%\centerline{HNC(1-0)}
\end{minipage}
\begin{minipage}{.25\textwidth}
\vspace{-8mm}
\includegraphics[width=\textwidth]{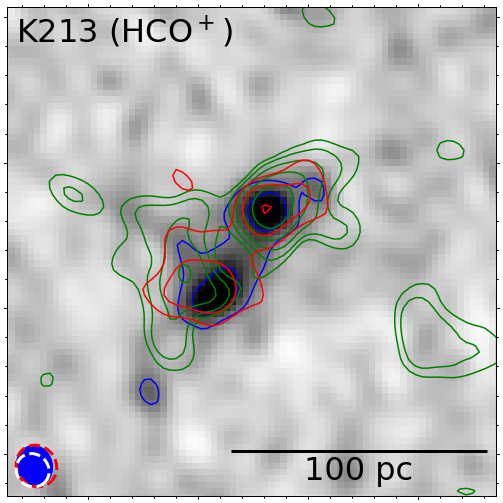}
%\centerline{HCO$^+$(1-0)}
\end{minipage}

{\bf Figure 1}, continued.
\end{figure*}

\begin{table*}
\caption{Observed sample\label{tab_sample}}
\begin{tabular}{llllll}
\hline
[KGF2015] & RA/Dec & 230 GHz cont. (SMA) &  & HCN(1-0) (NOEMA) & \\
 & (J2000) & synth. beam size ($''$) & rms (mJy) & synth. beam size ($''$) & rms (mK\,km\,s$^{-1}$)\\
\hline
 26 & 00:41:30.3 +41:04:55 & $4.5 \times 4.0$ & 0.19 & $4.4 \times 4.1$ & 83 \\
191 & 00:44:01.6 +41:49:09 & $4.2 \times 3.3$ & 0.40 & $4.4 \times 4.1$ & 96 \\
213 & 00:42:13.5 +40:51:21 & $4.3 \times 3.7$ & 0.31 & $4.4 \times 4.1$ & 88 \\
\hline
136 & 00:44:30.1 +41:51:48 & $5.6 \times 4.7$ & 0.20 & $4.4 \times 4.2$ & 57 \\
176 & 00:45:00.3 +41:28:36 & $8.1 \times 5.1$ & 0.24 & $4.4 \times 4.1$ & 74 \\
190 & 00:44:30.6 +41:56:34 & $4.9 \times 4.2$ & 0.17 & $4.5 \times 4.2$ & 47 \\
\hline
\end{tabular}
\end{table*}

%%%%%%%%%%%%%%%%%%%%%%%%%%%%%%%%%%%%%%%%%%%%%%%%%%%%%%%%%

\section{Results}\label{sec:res}

For a general overview of our results, we first present integrated emission maps of individual transitions as well as the continuum. We then next address the question of what the comparison of the 3~mm and 1~mm continuum emission is telling us about the emission mechanism before discussing the molecular emission, with a focus on HCN.

\subsection{Integrated emission maps}

Maps showing the extent of the HCN(1-0), HCO$^+$(1-0), HNC(1-0), and $^{13}$CO(1-0) emission in the targeted GMCs are shown in Figure~\ref{fig_maps}. The three images in each row separately summarize the observations for each individual source. 
These transitions have all been observed simultaneously by NOEMA at very similar nominal sensitivity and at identical spatial sensitivity. Any differences in the spatial extent of emission in different transitions thus is not due to spatial filtering. This observation extends to comparisons with our SMA data, which have similar $(u,v)$ coverage and hence spatial sensitivity. Note that we did not detect significant emission from HCN, HCO$^+$ or HNC from one source, K190, and it is  not included in the figure or later analysis of the line observations.

Emission from $^{13}$CO(1-0) is found to be detected over the largest projected area in each source.\footnote{For a comparison of simultaneously obtained $^{12}$CO(2-1) and $^{13}$CO(2-1) data, obtained with the SMA, see \citet{via21}.}  The HCN(1-0) and HCO$^+$(1-0) emission is detected over very similar areas, indicating that the emission from these molecules is approximately co-spatial at the sensitivity of our observations.
The emission in both molecular transitions also appears to be broadly co-spatial with the continuum emission, although the HCN emission can extend over slightly larger areas than the continuum emission. However, the HNC(1-0) emission is systematically detected over smaller areas. This is most striking in sources K136 and K213, where within the area that emits HCN(1-0) and HCO$^+$(1-0) the emission of HNC(1-0) splits up into smaller, distinct sources. This indicates that the emission of HNC(1-0) is either intrinsically weaker than the emission in HCN(1-0) and HCO$^+$(1-0), or it is confined to smaller volumes, or both. This finding is compatible with those of our millimeter-wavelength survey of largely starless cores in the Pipe Nebula \citep{for14}, where HNC was detected in fewer cores that HCN or HCO$^+$.
On larger (kpc) scales, this finding is also compatible with the ratios, reported in \citet{jim19}, of stacked integrated intensities for HNC/HCN of $<<1$, also indicating weaker HNC than HCN emission.

\begin{figure}
\includegraphics[width=\columnwidth]{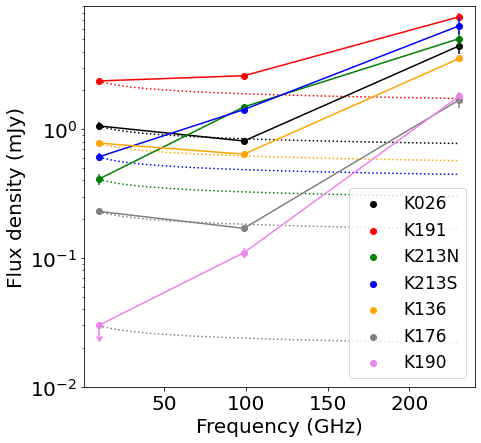}
\caption{Integrated flux densities of the observed sample, with continuum fluxes from VLA (10~GHz), NOEMA (99~GHz), and SMA (230 GHz) observations. For the sources with the strongest free-free continuum contamination in the millimeter range, dashed lines indicate extrapolations of the VLA measurements, assuming $S_\nu\propto \nu^{-0.1}$ for free-free emission.
\label{fig_rseds}}
\end{figure}

\subsection{The cm, 3~mm, and 1~mm continuum: free-free contamination}

The combination of our SMA, NOEMA, and VLA observations allows us to assess the nature of the observed emission, particularly where H\,{\sc ii} regions are present. While we would expect the thermal (greybody) dust emission in the millimeter and submillimeter wavelength range, any detections of H\,{\sc ii} regions in the centimeter wavelength range would be due to free-free emission, which due to its power-law behavior could contaminate the fluxes measured in the millimeter wavelength range.

While a full inventory of X-band (8-12~GHz) centimeter-wavelength detections toward the {\it Herschel} GMAs in M\,31 will be published by Toomey et al. ({\it in prep.}), we here use their data to assess any free-free contamination in the 3~mm (NOEMA) and 1~mm (SMA) bands for the present NOEMA sample. Assuming that the entire centimetric emission is from free-free processes in H\,{\sc ii} regions, it is possible to estimate the corresponding contamination at wavelengths of 3~mm and 1~mm, assuming and extrapolating $S_\nu\propto \nu^{-0.1}$ for free-free emission. 

These measurements are summarized in Figure~\ref{fig_rseds} and in Table~\ref{tab_rseds}, where the observed fluxes and free-free contamination fractions are listed at each wavelength.
Figure~\ref{fig_rseds} shows that the free-free emission completely dominates the flux in the 3~mm continuum in more than half of the sample (i.e., sources K26, K136, \& K176), where the 3~mm emission can be entirely explained as free-free emission. At 3~mm the average free-free contamination for the entire sample is 71 $\pm$ 35\%.  At 1~mm (230~GHz) the average free-free contamination is 13 $\pm$ 7 \%. 
All continuum detections at 1~mm  are clearly dominated by thermal dust emission. This is a key factor for using the dust continuum as a measure of dust and cloud mass when calculating mass-to-light ratios ($\alpha$) for molecular emission in the next section.

The degree of free-free contamination is not obvious from differences in the continuum maps at the two wavelengths, although in a few cases the 3~mm emission is clearly offset from the 1~mm continuum emission. While the 1~mm continuum is dominated by thermal dust emission, it may still be preferentially detecting emission from warmer dust next to H\,{\sc ii} regions.

\begin{table*} % {llllll}
\centering
\caption{Radio properties\label{tab_rseds}}
\begin{tabular}{llllll}
\hline
[KGF2015] & $S_{10~{\rm GHz}}$ & $S_{99~{\rm GHz}}$ & $S_{230~{\rm GHz}}$ & $F_{ff}$(99~GHz) & $F_{ff}$(230~GHz)\\
 & (mJy) & (mJy) & (mJy) & (\%) & (\%)\\
\hline
 26  & 1.06$\pm$0.04	    & 0.81$\pm$0.03 & 4.44$\pm$0.60 & 100   & 17 \\
191  & 2.37$\pm$0.03	    & 2.60$\pm$0.03 & 7.46$\pm$0.57 &  72   & 23 \\
213N & 0.41$\pm$0.04	    & 1.49$\pm$0.05 & 5.05$\pm$0.78 &  22   &  6 \\
213S & 0.61$\pm$0.04	    & 1.42$\pm$0.05 & 6.35$\pm$0.91 &  34   &  7 \\
\hline
136  & 0.78$\pm$0.03	    & 0.64$\pm$0.02 & 3.54$\pm$0.20 &  97   & 16 \\
176  & 0.23$\pm$0.01	    & 0.17$\pm$0.01 & 1.69$\pm$0.23 & 100   & 10 \\
190  & $<$0.03\,(3$\sigma$) & 0.11$\pm$0.01 & 1.80$\pm$0.16 & $<22$ & $<1$\\
\hline
\end{tabular}

{Flux errors are from Gaussian fitting and reflect source geometry. The estimated free-free contamination fraction in percent is shown in columns 5 and 6.}
\end{table*}

%%%%%%%%%%%%%%%%%%%%%%%%%%%%%%%%%%%%%%%%%%%%%%%%%%%%%%%%%%%

\section{Analysis and Discussion} \label{sec:ana}

Based on the dust continuum detections of individual clouds at 230~GHz, we can expand our previous work on direct measurements of $\alpha_{\rm CO}$ to the traditional dense-gas tracers HCN and HCO$^+$. As discussed above, their detections appear to be largely cospatial with that of the continuum emission. Thereafter, we intercompare the HCN, HCO$^+$, and HNC emission.

\subsection{Assessing HCN as a high-density tracer}

Because of the much higher dipole moment of HCN compared to CO and the fact that early surveys of Milky Way GMCs found HCN emission to be considerably less extended (e.g., R$_{HCN}$  $\sim$ 1 pc; \citealp{wue05}) in clouds than CO emission (e.g., R$_{CO}$ $\sim$ 30-100 pc; \citealp{ric16,miv17}), HCN has been long regarded as a tracer of the higher density (n(H$_2$) $>$ 10$^4$ cm$^{-3}$) gas in molecular clouds (e.g., \citealp{got75,sol92,jac96,hel97,eva99,gas04,wue05,jim19}, etc.). 
However, recent deep and more sensitive mapping observations of nearby molecular clouds has found HCN emission to be considerably more extended arising in regions with extinctions below those (e.g., A$_K$ $\geq$ 0.8 magnitudes)  associated with active star formation and gas densities  $\gtrsim$ 10$^4$ cm$^{-3}$ (e.g., \citealp{pet17,shi17,kau17,eva20,taf21,dal23}).  
This suggests that a significant fraction ($\sim$ 30-60 \%)  of the HCN emission from a cloud may originate in regions with densities n(H$_2$) $\sim$ 10$^3$ cm$^{-3}$, a value significantly below both the critical density and the densities most closely associated with active star formation. 
For example, in their exhaustive and deep survey of the Perseus cloud \citet{dal23} detected HCN emission over $\sim$50\% of the $^{12}$CO($J=2-1$) emitting area of that cloud.

In the GMCs of M31 \citet{via21} found that the areal extent of $^{13}$CO  emission was typically 70\% of that of $^{12}$CO.  Consequently, if our NOEMA observations in M31 had the same absolute sensitivity that Dame \& Lada had for Perseus, we would expect to see HCN emission extended over a significant fraction of the $^{13}$CO emitting area.  
However, examination of Figure 2 clearly shows that this is not the case. This is not surprising since M31 is roughly 2000 times more distant than Perseus. Our NOEMA observations appear only sensitive enough to be detecting the tips of the HCN ``icebergs''. Because HCN emission has been found to be also a good tracer of total mass column density \citep{taf21,dal23} we are likely detecting only the highest column density gas in M31 GMCs with our current HCN observations. The same is likely true for the optically thin 230~GHz dust continuum emission, and the near-coincidence of the two tracers in M31 suggests that these measurements are both effectively and independently tracing the same material, which happens to be the highest column density gas in the clouds. However, it is difficult to quantitatively determine the corresponding column and volume densities of this gas with any certainty. Analogy with the Perseus cloud would suggest that the emission we observed is dominated by  the kind of  high densities  (i.e., A$_K$ $\gtrsim$ 0.8 magnitudes, n(H$_2$) $\gtrsim$ 10$^4$ cm$^{-3}$ ) associated with imminent or active star formation in the Milky Way. However, we caution that comparisons with the Perseus cloud are limited by the fact that it remains to be determined if the Perseus cloud or any single molecular cloud can be considered as representative of GMCs in general.

\subsection{Determination of $\alpha^\prime_{\rm HCN}$ and $\alpha^\prime_{\rm HCO^+}$}

Following our earlier work \citep{for20,via21} we calculate $\alpha'$, the mass-to-luminosity ratio for the dust:
$$\alpha' = {M_{dust}\over{L}}$$
where $M_{dust}$ is the dust mass in units of solar mass (M$_\odot$) and $L$ is the luminosity of the molecular transition under consideration, expressed in units of K km s$^{-1}$ pc$^2$. Although this ratio converts the HCN or HCO$^+$ emission to a  dust mass, not a total mass, $\alpha'$ can be scaled to a more standard conversion factor for the total mass ($\alpha$) by assuming a gas-to-dust ratio.
Having confirmed that the continuum emission at 230~GHz is indeed dominated by thermal dust emission, we can proceed to a relatively direct measurement of $\alpha'_{\rm HCN}$ and $\alpha'_{\rm HCO^+}$. However, we first need to correct the 230~GHz flux ($S_\nu$) for the estimated free-free emission before then converting the continuum flux to a dust mass employing  the relation $M_{dust} = S_\nu d^2/(\kappa_\nu B_\nu(T_d))$ using the parameters assumed in our earlier studies (i.e., $\kappa_\nu = 0.0425\ {\rm m^2kg^{-1}}$, $T_d = 20\ K$, $d=780\ {\rm kpc}$). 

While, in contrast to our earlier determinations of $\alpha^\prime_{\rm CO}$ at 230~GHz, the HCN(1-0), HCO$^+$(1-0) and 230~GHz continuum data have not been obtained simultaneously, with identical $(u,v)$ coverage, their $(u,v)$ coverage and resulting resolution are very similar (see Table~\ref{tab_sample} and Figure 1). While in a few cases, the resolution is practically identical, the synthesized beam area otherwise differs by values ranging from 10\% to a factor of 2 (in the case of K176). We do not, however, consider spatially resolved $\alpha_{\rm HCN}$ and $\alpha_{\rm HCO^+}$ measurements, but obtain the values of $\alpha$ using only integrated measures of the luminosities and masses as described below.

\begin{figure*}
\centering
\begin{minipage}{.3\textwidth}
\includegraphics[width=\textwidth]{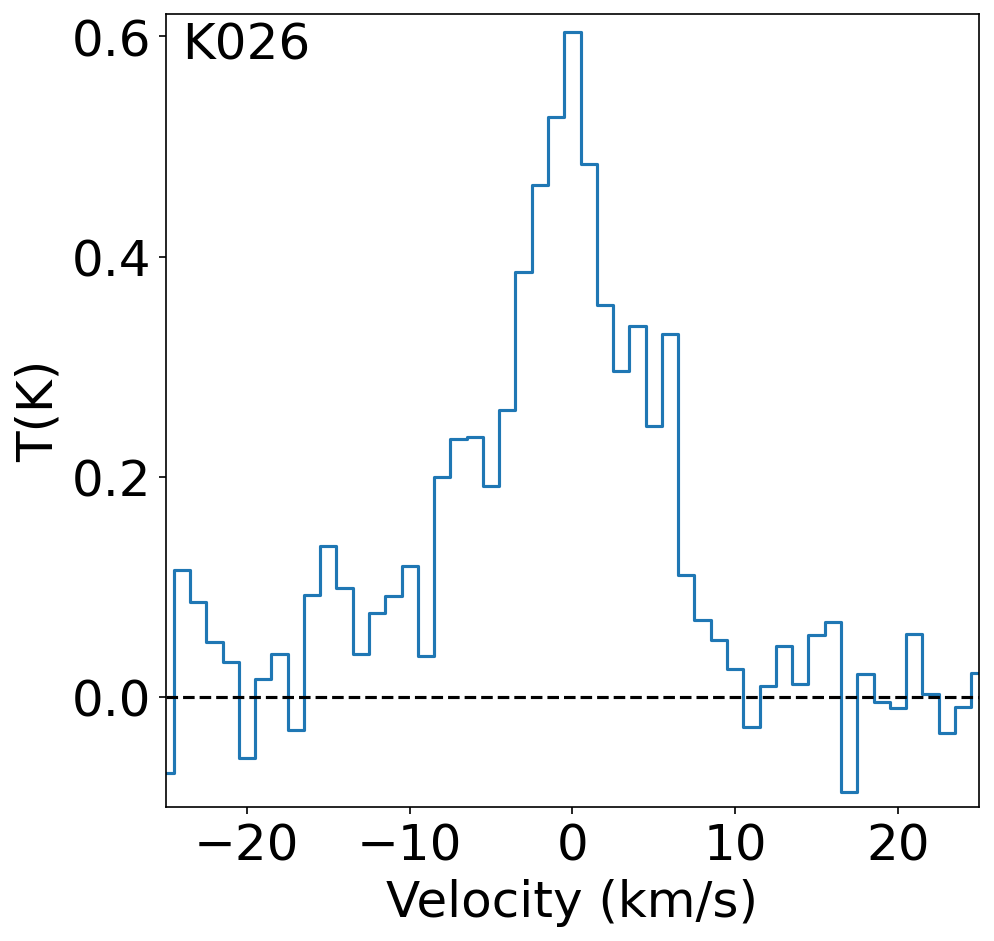}
%\centerline{K026}
\end{minipage}
\begin{minipage}{.3\textwidth}
\includegraphics[width=\textwidth]{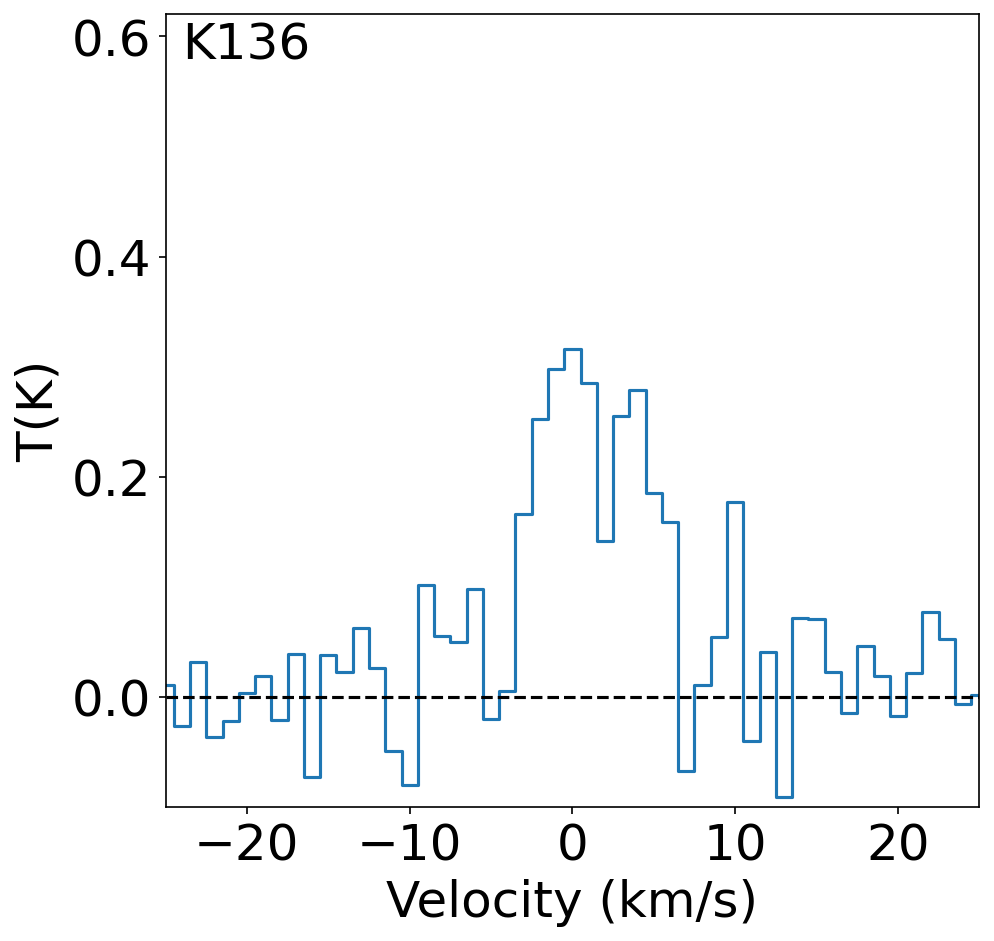}
%\centerline{K136}
\end{minipage}
\begin{minipage}{.3\textwidth}
\includegraphics[width=\textwidth]{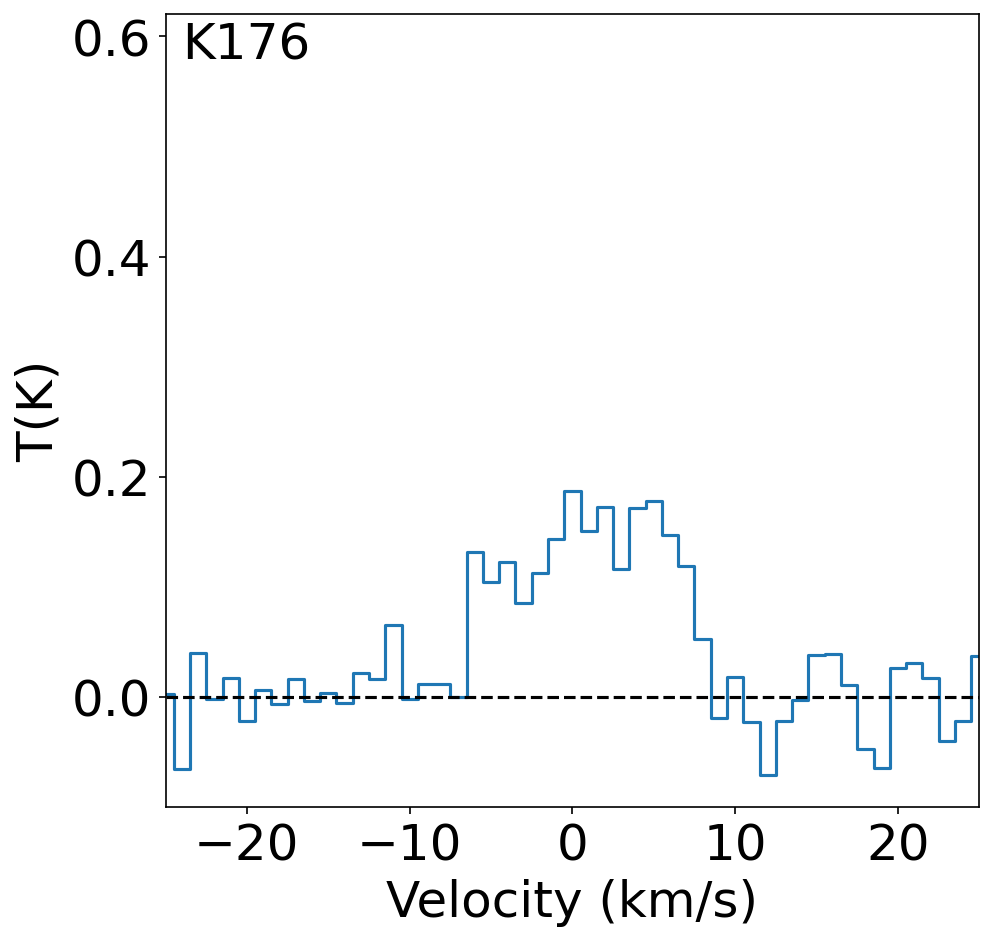}
%\centerline{K176}
\end{minipage}

\begin{minipage}{.3\textwidth}
\includegraphics[width=\textwidth]{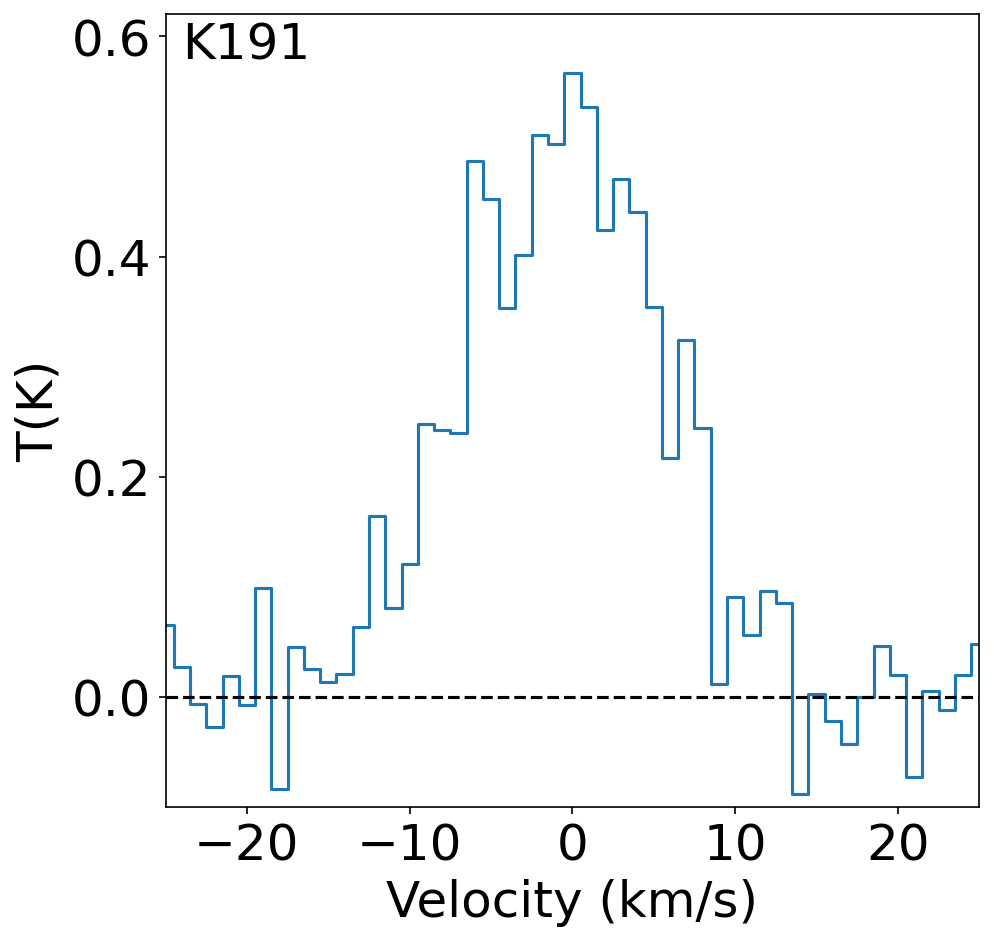}
%\centerline{K191}
\end{minipage}
\begin{minipage}{.3\textwidth}
\includegraphics[width=\textwidth]{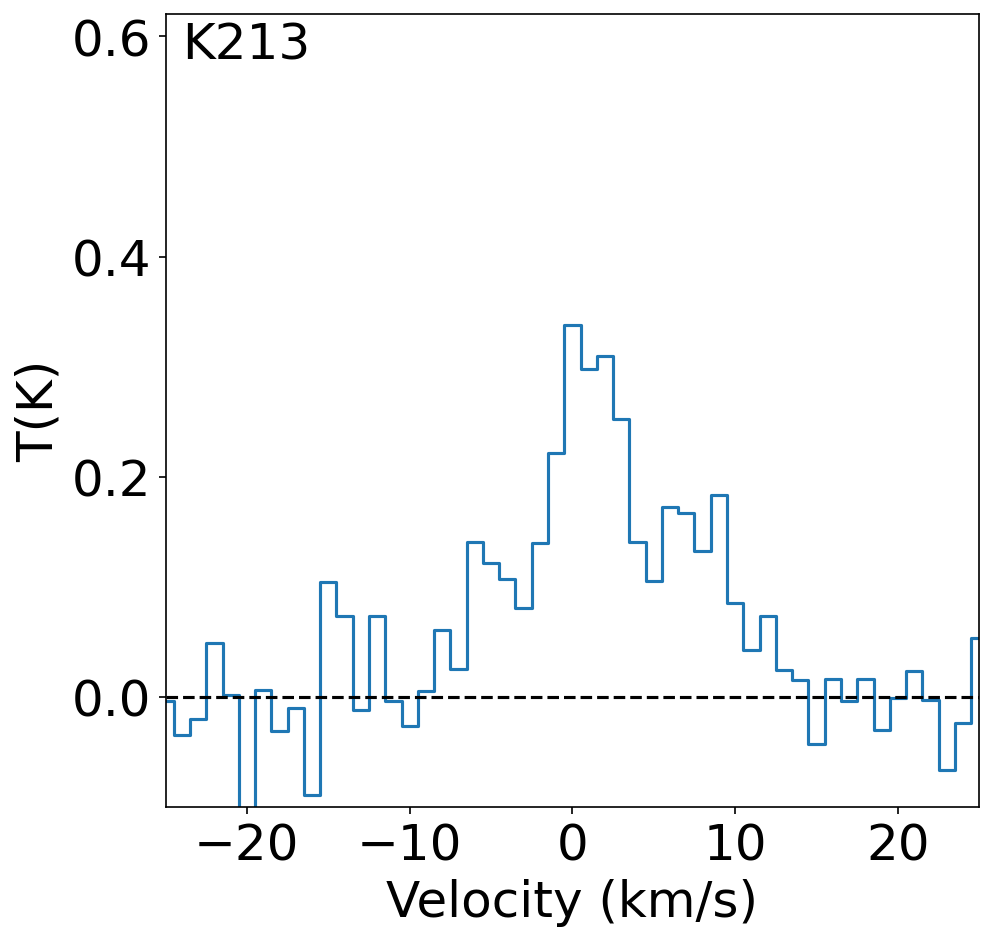}
%\centerline{K213}
\end{minipage}

\caption{HCN spectra extracted from the 3$\sigma$ contours of the corresponding moment~0 maps. The brightness temperature scale has been obtained from the native flux densities together with the respective synthesized beam size.
\label{fig_spec}}
\end{figure*}

Although in the present analysis we generally follow the procedure described in \citet{for20}, we do not require the detection of both HCN (or HCO$^+$) and the dust continuum in a joint mask, but instead we separately evaluate the molecular luminosities  and the dust masses each within their respective 3$\sigma$ contours. This maximizes the signal in a situation where two types of emission are already largely coincident. The spatially integrated HCN spectra within the 3$\sigma$ contours that were used to calculate the line luminosities are shown in Figure~\ref{fig_spec}. The lines are clearly detected, with hints of hyperfine structure, which however cannot be extracted with sufficient accuracy.  We have additionally extracted the HCN and HCO$^+$ luminosities within the 3$\sigma$ contours of $^{13}$CO in an attempt to also capture faint extended HCN and HCO$^+$, including by extracting HCN spectra in-between the nominal 3$\sigma$ contours of $^{13}$CO and HCN. As a result, we found only insignificant changes to the total line luminosities and hence $\alpha^\prime$ values, indicating that any faint extended emission has no significant impact on our measurements, most likely since any such emission is below our sensitivity cut-off in M31. However, even though no additional emission is added by integrating within the $^{13}$CO contours, the uncertainty of the derived values for $\alpha^\prime_{\rm HCN}$ would still increase significantly due to the larger projected area covered.

We present the resulting $\alpha^\prime_{\rm HCN}$ and $\alpha^\prime_{\rm HCO^+}$ measurements in Table~\ref{tab_hcnalpha}.
The average value of $\alpha^\prime_{\rm HCN}$ is 1.1 $\pm$ 0.7 M$_\odot$(K km/s pc$^2$)$^{-1}$. The dispersion in the average is relatively large, mostly due to source K213 which seems to be an outlier both in the value of $\alpha_{\rm HCN}$ and its uncertainty. The average without K213 is 0.80 $\pm$ 0.21 M$_\odot$(K km/s pc$^2$)$^{-1}$ which indicates a relatively small variation ($\sim$ 25\%) in the cloud to cloud value of $\alpha_{\rm HCN}$ in these clouds. For a gas-to-dust ratio similar to that of the Milky Way, this measurement corresponds to a value of $\alpha_{\rm HCN}\sim 109\pm23$~$M_\odot$\,(K\,km\,s$^{-1}$\,pc$^{2}$)$^{-1}$. Although suggestive, the sample is too small at this time to be able to draw any definitive conclusions regarding the general uniformity of the dense gas conversion factor in the M31 cloud population. 
For K190, with undetected HCN, we obtain a nominal limit of $\alpha^\prime_{\rm HCN}>1.3$~M$_\odot$(K km/s pc$^2$)$^{-1}$, when conservatively assuming the 3$\sigma$ limit of HCN within the $^{13}$CO 3$\sigma$ contour. Source K190 does not constitute an outlier when compared with the other sources.

Ideally, we would compare our M31 measurements with those of  nearby clouds in the Milky Way, but with very few exceptions, no local clouds have been observed in either HCN or HCO$^+$ in their entirety. The recent HCN observations of the Perseus molecular cloud by \citet{dal23} offer such a rare exception for the most nearby clouds.
We can compare the values of $\alpha'_{\rm HCN}$ found here with that reported for the Perseus  cloud. As mentioned earlier, \citet{dal23} used a complete survey of HCN  to obtain a robust measurement of the cloud's total HCN luminosity and directly compared it with the dense gas mass of the cloud derived from detailed observations of the dust \citep{zar16}.  This yielded a direct measurement of the dense gas $\alpha_{\rm HCN}$. Assuming a gas-to-dust ratio of 136, characteristic of the Milky Way, their measurement corresponds to $\alpha'_{HCN}$ $=$ 0.68 M$_\odot$(K km/s pc$^2$)$^{-1}$. 

We cannot yet make a direct comparison of the Dame \& Lada result with the values in Table 3 because in the former study the $\alpha^\prime_{\rm HCN}$ corresponds to a dense gas conversion factor where the HCN luminosity corresponds to the HCN luminosity from the entire cloud and the mass is the mass of dense gas (i.e., A$_K$ $\geq$ 0.8 magnitudes)  of the cloud, whereas in Table 3 the HCN luminosity only corresponds to the area of each cloud with detectable emission lines and not to that of the whole cloud and the mass corresponds to the mass of the HCN emitting area of the cloud. However, we can attempt to determine the HCN luminosity that might originate in the rest of a cloud by spatially integrating all the pixels within the $^{13}$CO emission area of each cloud that do not contain already detectable HCN emission according to the moment~0 map.\footnote{ Use of the $^{13}$CO emitting area in this context is reasonable in light of the fact that the deep survey of the Perseus cloud by \citet{dal23} found the HCN emitting area of that cloud to be about 50\% of the total $^{12}$CO emitting area.}  
Performing this exercise we did not detect any additional HCN emission in any of the clouds. However, from the 3$\sigma$ upper limits of the moment~0 maps we can estimate  upper limits to the total HCN luminosity for each source, assuming that an area with a size of the $^{13}$CO-emitting region is filled with HCN emission at the 3$\sigma$ level. From this exercise  we found the upper limits to the total HCN luminosity to be on average a factor of 2 above the measured values in Table 3. Assuming the dust masses derived in Table 3 correspond to dense gas masses we use these upper limits on L(HCN) to calculate lower limits to $\alpha^\prime_{\rm HCN}$  and find them to  range from $>$0.26 to $>$0.84 with an average of $\alpha^\prime_{\rm HCN}$ $>$ 0.5 $\pm$ 0.2 which is  74\% of the Dame \& Lada value for the Perseus cloud. This corresponds to $\alpha_{\rm HCN}$ $>$ 68 for a gas-to-dust ratio of 136.   Given that these are lower limits, it appears that, from comparison with the Dame \& Lada result, our measurements are consistent with the notion that the  values of $\alpha^\prime_{\rm HCN}$ derived from our NOEMA observations for M31 GMCs correspond to dense gas conversion factors where the 230~GHz dust continuum emission measured by the SMA and the HCN emission measured by NOEMA in these cases both trace the dense gas in these clouds. Similar considerations would apply to our HCO$^+$ observations.

A second relevant example is that of the W49 cloud, a site of high-mass star formation at a distance of 11~kpc, reported by \citet{bar20}. For the conversion of HCN luminosity to dense gas at $A_V>8$~mag, this study found $\alpha_{\rm HCN} = 30$~$M_\odot$(K km/s pc$^2$)$^{-1}$, corresponding to $\alpha^\prime_{\rm HCN} = 0.22$~$M_\odot$(K km/s pc$^2$)$^{-1}$, where we again assume a standard gas-to-dust ratio of 136, as in the Perseus case above. This value is slightly below our results in M31 but still broadly compatible.

\begin{table*} % {llllll}
\caption{HCN luminosity to dust mass comparison\label{tab_hcnalpha}}
\begin{tabular}{llllll}
\hline
[KGF2015] & $L_{\rm HCN}$ & $L_{\rm HCO^+}$ & $M_{\rm dustcorr}$ & $\alpha^\prime_{\rm HCN}$ & $\alpha^\prime_{\rm HCO^+}$\\
 & (K\,km\,s$^{-1}$\,pc$^2$) & (K\,km\,s$^{-1}$\,pc$^2$) & ($M_\odot$) & ($M_\odot$\,(K\,km\,s$^{-1}$\,pc$^2$)$^{-1}$) & ($M_\odot$\,(K\,km\,s$^{-1}$\,pc$^2$)$^{-1}$)\\
\hline
 26  & 1225$\pm$131 & 1221$\pm$271 & 654$\pm$73   & 0.53$\pm$0.19 & 0.54$\pm$0.18\\
136  &  477$\pm$172 & 436$\pm$132  & 487$\pm$67   & 1.02$\pm$0.51 & 1.12$\pm$0.49\\
176  &  386$\pm$106 & 595$\pm$157 &  347$\pm$60   & 0.90$\pm$0.40 & 0.58$\pm$0.25\\
191  & 1886$\pm$104 & 1413$\pm$310 & 1420$\pm$155  & 0.75$\pm$0.22 & 1.00$\pm$0.33\\
213  &  820$\pm$88  & 1097$\pm$483 & 1961$\pm$184  & 2.39$\pm$1.10 & 1.79$\pm$0.96\\
\hline
\end{tabular}
\end{table*}

The methodology of \citet{dal23} differs slightly from the one used here in that the former study calculated cloud masses using extinction calibrated emission maps from the {\sl Herschel} satellite \citep{zar16} whereas here we obtained cloud masses from a standard procedure using continuum observations made at a single frequency (230 GHz). It is of some interest to evaluate the extent to which these different approaches might affect the derived cloud masses and thus $\alpha$s. To perform such a consistency check  we obtained a dust mass for the entire Perseus cloud from the {\it Planck} 217~GHz continuum map of the region, assuming the same parameters of dust temperature and opacity used for the procedure adopted here for the M31 clouds (see above and \citealp{for20,via21}). 
To define the cloud mass, we use a contour corresponding to the outer boundary of the cloud at $A_K=0.1$~mag, based on the conversion of the matching {\it Planck} $\tau_{\rm 353~GHz}$ map, as discussed in \citet{lel22}. We find a total dust mass of 193~$M_\odot$, which leads to a conversion factor of $\alpha^\prime_{\rm HCN} = 3.5$~$M_\odot$\,(K\,km\,s$^{-1}$\,pc$^{2}$)$^{-1}$, or  $\alpha_{\rm HCN} = 470$~$M_\odot$\,(K\,km\,s$^{-1}$\,pc$^{2}$)$^{-1}$ if we assume a gas-to-dust ratio of 136. This value does not correspond to the dense gas $\alpha_{\rm HCN}$ since it applies to the total cloud mass. This is in excellent agreement with the analogous value (500) derived for the whole cloud based on extinction mapping in \citet{dal23}. 
No such comparison is available for HCO$^+$.

\subsubsection{Comparison of HCN and HCO$^+$ emission}

It is a direct corollary of the determination of very similar values for $\alpha^\prime_{\rm HCN}$ and $\alpha^\prime_{\rm HCO^+}$ as well as the near-cospatial emission that we find a ratio of the HCN and HCO$^+$ luminosities of close to unity. While both molecules are collisionally excited, one of them is an ion, which means that the ratio will not only be affected by the ratio of critical densities but also the ionization. 
When compared to HCN(1-0), the critical density of HCO$^+$(1-0) is nearly an order of magnitude lower. The HCN/HCO$^+$ ratio thus depends on gas density, with HCO$^+$ favored at lower densities. In high-density gas, the abundance of HCO$^+$ decreases due to the faster recombination with electrons (e.g., \citealp{pap07}). These two effects suggest that HCN should be brighter than HCO$^+$ in high-density gas. Perhaps more importantly, differences between the two molecules could also reflect differences in abundance, since the abundance of HCO$^+$(1-0) could be enhanced by ionization in H\,{\sc ii} regions when compared with more quiescent regions.

As before, we extract the HCN(1-0) and HCO$^+$(1-0) intensities in a joint mask, where both lines are detected above S/N$>$3. Contrary to the comparison between HCN and HNC (see below), all sources show very similar morphologies in both transitions, including both components of K213. In all cases, the intensity ratio is compatible with a value of unity, with only minor and insignificant differences, as listed in Table~\ref{tab_hcophcn}. 

Our range of observed ratios is compatible with the values reported by \citet{bro05} using single-dish data. We have one source in common with \citet{bro05}, since K026 is coincident with source D. Our ratio of $I({\rm HCO^{+}})/I({\rm HCN}) = 1.0 \pm 0.3$ is compatible with that found by \citet{bro05}, which is $I({\rm HCO^{+}})/I({\rm HCN}) = 1.9 \pm 0.8$. More generally, they detect both lines in eight clouds and find that the HCO$^+$ emission is on average 20\% stronger than the HCN emission, but with low significance. In the Large Magellanic Cloud, HCO$^+$/HCN flux ratios of 0.5 to 0.8 have been reported (e.g., \citealt{gal20} and references therein).

Our finding of a flux ratio of HCO$^{+}$/HCN$\sim 1$ with cospatial emission either means that very dense gas is observed, such that the difference in critical density of the transitions no longer matters, or the HCO$^+$ emission could be enhanced due to ionization in H\,{\sc ii} regions. We note that the linewidths of HCN and HCO$^+$ are similar as well, supporting the idea that both trace the same gas, but the linewidths cannot be measured at high confidence due to the S/N of the spectra, particularly where the hyperfine structure of HCN(1-0) is concerned.

\begin{table}
\caption{$I({\rm HCO^{+}})/I({\rm HCN})$ intensity ratios\label{tab_hcophcn}}
\begin{tabular}{ll}
\hline
[KGF2015] & $I({\rm HCO^{+}})/I({\rm HCN})$\\
\hline
 26  & 1.0$\pm$0.3 \\
136  & 1.0$\pm$0.2 \\
176  & 1.2$\pm$0.2 \\
191  & 0.8$\pm$0.2 \\
213N & 1.0$\pm$0.2 \\
213S & 1.2$\pm$0.2 \\
\hline
\end{tabular}
\end{table}

\subsection{Comparison of HCN and HNC emission}

While the HCN emission is not always cospatial with HNC, as described above, this leaves us with HCN/HNC as another relevant parameter in describing the physical conditions in the observed GMCs.

The line ratio HCN/HNC is $\approx$1 in cold, dark clouds, but it has been shown to vary by a factor of $>10$, depending on $T_{\rm kin}$ and $n({\rm H}_2)$, since HNC is converted into HCN at higher temperatures \citep{gol86,sch92}. More recently, based on a comparison with kinetic temperatures from NH$_3$ assumed to be cospatial, and on much smaller scales ($<0.1$~pc) than those considered in this work, the intensity ratio HCN(1-0)/HNC(1-0) has been proposed as a proxy measurement for ambient temperature \citep{hac20}, assuming that the emission is cospatial, including with NH$_3$, which is used for kinetic temperature calibration. This is particularly relevant since HCN and HNC transitions can often be observed simultaneously with the same observational setup, limiting the impact of observational systematics and uncertainties.

To obtain intensity ratios, we integrate the HCN(1-0) and HNC(1-0) emission in joint masks where both lines are detected above S/N$>$3. This analysis excludes K190, where the lines were not detected, and it also excludes the two interesting cases of K136 and K213S, where the HCN and HNC emission clearly comes from very different areas, with only limited overlap (see Figure~\ref{fig_maps} above). The resulting intensity ratios for the four sources that show cospatial emission at the resolution of our measurements cluster around $I({\rm HCN})/I({\rm HNC})\approx 2.5$, with just one marginal outlier at a slightly higher value. The ratios are listed in Table~\ref{tab_hcnhnc}.

\begin{table}
\caption{$I({\rm HCN})/I({\rm HNC})$ intensity ratios\label{tab_hcnhnc}}
\begin{tabular}{ll}
\hline
[KGF2015] & $I({\rm HCN})/I({\rm HNC})$\\
\hline
 26  & 2.4$\pm$0.4 \\
176  & 2.5$\pm$0.3 \\
191  & 3.5$\pm$0.7 \\
213N & 2.5$\pm$0.3 \\
\hline
\end{tabular}
\end{table}

At face value, these measurements indicate a corresponding kinetic ammonia temperature of about 25~K, while the measurement for K191 may be as high as 35~K, but the difference in the measured ratios is not significant. Additionally, the relation reported by \citet{hac20} has scatter on a similar scale as our measurement uncertainties. For the regions where we detected cospatial HCN and HNC emission, one would thus expect corresponding ammonia kinetic temperatures between 20~K and 40~K. However, the fact that two of our six targets do show detections in both HCN and HNC that are not cospatial indicates that there may be no generally applicable conversion of the intensity ratio to temperature for these sources.

%%%%%%%%%%%%%%%%%%%%%%%%%%%%%%%%%%%%%%%%%%%%%%%%%%%%%%

\section{Summary} \label{sec:sum}

We report observations of HCN(1-0), HCO$^+$(1-0), HNC(1-0), $^{13}$CO(1-0)  and 100 GHz continuum emission obtained  toward six GMCs in M31 using NOEMA. These clouds were previously identified, resolved and imaged at 230 GHz in both CO(2-1) and continuum emission with the SMA. Results of our analysis of these observations are summarized as follows:

\begin{itemize}

 \item The maps of the five clouds that show HCN(1-0), HCO$^+$(1-0), and HNC(1-0) emission indicate that in each cloud the emission from these molecules is spatially resolved and overlapping with very similar spatial extents, while emission from $^{13}$CO(1-0) is observed over much larger areas. The continuum emission at both 100 and 230 GHz spatially overlaps with and is similarly extended as the HCN, HCO$^+$(1-0), and HNC(1-0) emission, consistent with the notion that the continuum observations trace the dense gas component of the GMCs.
 
\item In the five clouds with radio-detected H\,{\sc ii} regions, comparison of the continuum emission at 100 and 230 GHz with existing VLA observations at 10 GHz  shows that free-free emission accounts for 22-100\% of the continuum flux at 100 GHz, but only 6-23\% of the emission at 230 GHz demonstrating the need for continuum observations at the higher frequencies to obtain accurate measurements of thermal dust emission and consequently dust masses in such clouds.

\item We directly derive the dust mass conversion factor or mass-to-light ratio, $\alpha^\prime_{HCN}$, by comparing the total luminosity of HCN to the dust mass derived from the 230 GHz continuum observations corrected for the measured free-free contamination in the clouds. We find a mean of $<\alpha^\prime_{\rm HCN}> = 1.1\pm0.7$~$M_\odot$\,(K\,km\,s$^{-1}$\,pc$^{2}$)$^{-1}$, where the standard deviation is dominated by a single outlier. Removing the outlier cloud results in $<\alpha^\prime_{\rm HCN}> = 0.80\pm0.21$~$M_\odot$\,(K\,km\,s$^{-1}$\,pc$^{2}$)$^{-1}$. Under the assumption that the gas-to-dust ratio is similar to that of the Milky Way, this latter measurement corresponds to a value of $\alpha_{\rm HCN}\sim 
109\pm23$~$M_\odot$\,(K\,km\,s$^{-1}$\,pc$^{2}$)$^{-1}$.

\item While higher than most previously reported HCN conversion factors, our measurement of $\alpha_{\rm HCN}$ is compatible with the dense gas calibrated value (92 $M_\odot$\,(K\,km\,s$^{-1}$\,pc$^{2}$)$^{-1}$) recently derived by \citet{dal23} based on an analysis of complete HCN and dust extinction maps of the Perseus molecular cloud. The relatively close agreement of these measures could imply that the HCN gas conversion factors we derive here for M31 likely correspond to dense gas conversion factors, but this will have to be checked on a larger, more diverse sample of GMCs in both the Milky Way and M\,31.

\item Similar to our HCN determination, for HCO$^+$, we find a value of $\alpha^\prime_{\rm HCO^+} = 0.81 \pm 0.29$~$M_\odot$\,(K\,km\,s$^{-1}$\,pc$^2$)$^{-1}$. With the same assumptions for the gas-to-dust ratio as before, we obtain a value for $\alpha_{\rm HCO^+}\sim 110$~$M_\odot$\,(K\,km\,s$^{-1}$\,pc$^2$)$^{-1}$.
\item We find an enhanced ratio of HCO$^{+}$/HCN$\sim 1$, but it remains unclear whether this is due to high densities or an enhanced abundance of HCO$^{+}$ in the vicinity of H\,{\sc ii} regions.
\item Based on work by \citet{hac20} comparing HCN(1-0) to HNC(1-0) emission ratios with ammonia kinetic temperatures, we conclude from the emission ratios in our sample that the typical temperatures traced by our HCN  observations are $T\sim 25$~K, with an uncertainty of about 10~K.
\item When comparing HCN(1-0) and HCO$^+$(1-0) emission, we find the ratio to be compatible with unity in all cases. Our results are compatible with previous single-dish observations of M31 reported by \citet{bro05}.

\end{itemize}

\section*{Acknowledgements}

We thank the referee, Frank Bigiel, for constructive comments that helped to clarify the paper, and Mario Tafalla, for helpful discussions. This work is based on observations carried out under project numbers S20AX and W20BN with the IRAM NOEMA Interferometer. IRAM is supported by INSU/CNRS (France), MPG (Germany) and IGN (Spain).

%%%%%%%%%%%%%%%%%%%%%%%%%%%%%%%%%%%%%%%%%%%%%%%%%%
\section*{Data Availability}

This paper makes use of IRAM datasets S20AX and W20BN, with data available from the IRAM Science Data Archive after the proprietary period. Science products will be made available upon reasonable request to the authors.



\bibliographystyle{mnras}
\bibliography{m31}

%\begin{thebibliography}{99}
%\end{thebibliography}

%%%%%%%%%%%%%%%%%%%%%%%%%%%%%%%%%%%%%%%%%%%%%%%%%%

%%%%%%%%%%%%%%%%% APPENDICES %%%%%%%%%%%%%%%%%%%%%

%\appendix



%%%%%%%%%%%%%%%%%%%%%%%%%%%%%%%%%%%%%%%%%%%%%%%%%%

% Don't change these lines
\bsp	% typesetting comment
\label{lastpage}
\end{document}